\documentclass{ws-procs975x65}
\usepackage{graphicx}

\def\beq{\begin{equation}}
\def\eeq{\end{equation}}

\begin{document}

\title{Torsion-bar antenna: a ground-based mid-frequency and low-frequency gravitational wave detector}
\author{Tomofumi Shimoda$^*$, Satoru Takano, Ching Pin Ooi, Naoki Aritomi, Yuta Michimura and Masaki Ando}

\address{Department of Physics, University of Tokyo,\\
Tokyo, 113-0033, Japan\\
$^*$E-mail: shimoda@granite.phys.s.u-tokyo.ac.jp}

\author{Ayaka Shoda} 

\address{National Astronomical Observatory of Japan, Osawa 2-21-1,\\
Mitaka, Tokyo 181-8588, Japan}

\begin{abstract}
Expanding the observational frequency of gravitational waves is important for the future of astronomy. 
Torsion-Bar Antenna (TOBA) is a mid-frequency and low-frequency gravitational wave detector using a torsion pendulum.
The low resonant frequency of the rotational mode of the torsion pendulum enables ground-based observations.
The overview of TOBA, including the past and present status of the prototype development is summarized in this paper.

\end{abstract}


\bodymatter


\section{Introduction}
Recent detections of gravitational waves (GWs) by Advanced LIGO and Advanced Virgo \cite{GW150914, GW170817, GW_O2}, and their electromagnetic follow-up observations have opened an era of multi-messenger astronomy.
GWs successfully provided unique information about ten binary black holes and a binary neutron star.
The number of detection will increase as the sensitivities of the detectors improve, leading to a better understanding of the universe.
At the same time, several proposed space-based GW detectors such as LISA \cite{LISA}, DECIGO \cite{DECIGO}, AIGSO \cite{AIGSO} and AMIGO \cite{AMIGO} will enlarge the observational frequency band to below 1 Hz.
With such mid-frequency (100 mHz to 10 Hz) and low-frequency (100 nHz to 100 mHz) GWs, information on massive ($M > 10^3 M_{\rm sun}$) black holes is expected to be available.

In addition, some ground-based approaches to observe mid-frequency and low-frequency GWs have been proposed.
One of which is Torsion-Bar Antenna (TOBA) \cite{TOBA0}, which uses torsion pendulums to observe GWs.
Since a mechanical oscillator acts as a free-falling mass above its resonant frequency, the low resonant frequency of the torsion pendulum enables GW observation around few tens of millihertz.
Its target sensitivity is $10^{-19} /\sqrt{\rm Hz}$ at 0.1 Hz, which allows us to detect intermediate mass black holes (IMBHs) at cosmological distances.
While the space-based detectors are expected to have better sensitivities at similar frequencies, the ground-based configurations are of lower cost and have better accessibility.
Other proposals of ground-based detectors include SOGRO \cite{SOGRO} and MIGA \cite{MIGA}, which use superconducting levitated masses and atomic interferometers, respectively.

In this paper, we summarize the overview of TOBA.
The target sensitivity and the observation ability are shown in Sec. \ref{sec:target}.
Then the results of prototype development and the upcoming plan are explained in Sec. \ref{sec:prototype}.
Sec. \ref{sec:phase3} introduces the important next step, phase-III TOBA.

\section{Principle and target of TOBA}\label{sec:target}

\def\figsubcap#1{\par\noindent\centering\footnotesize(#1)}
\begin{figure}[t]%
\begin{center}
	\parbox{2.8in}{\includegraphics[width=2.8in]{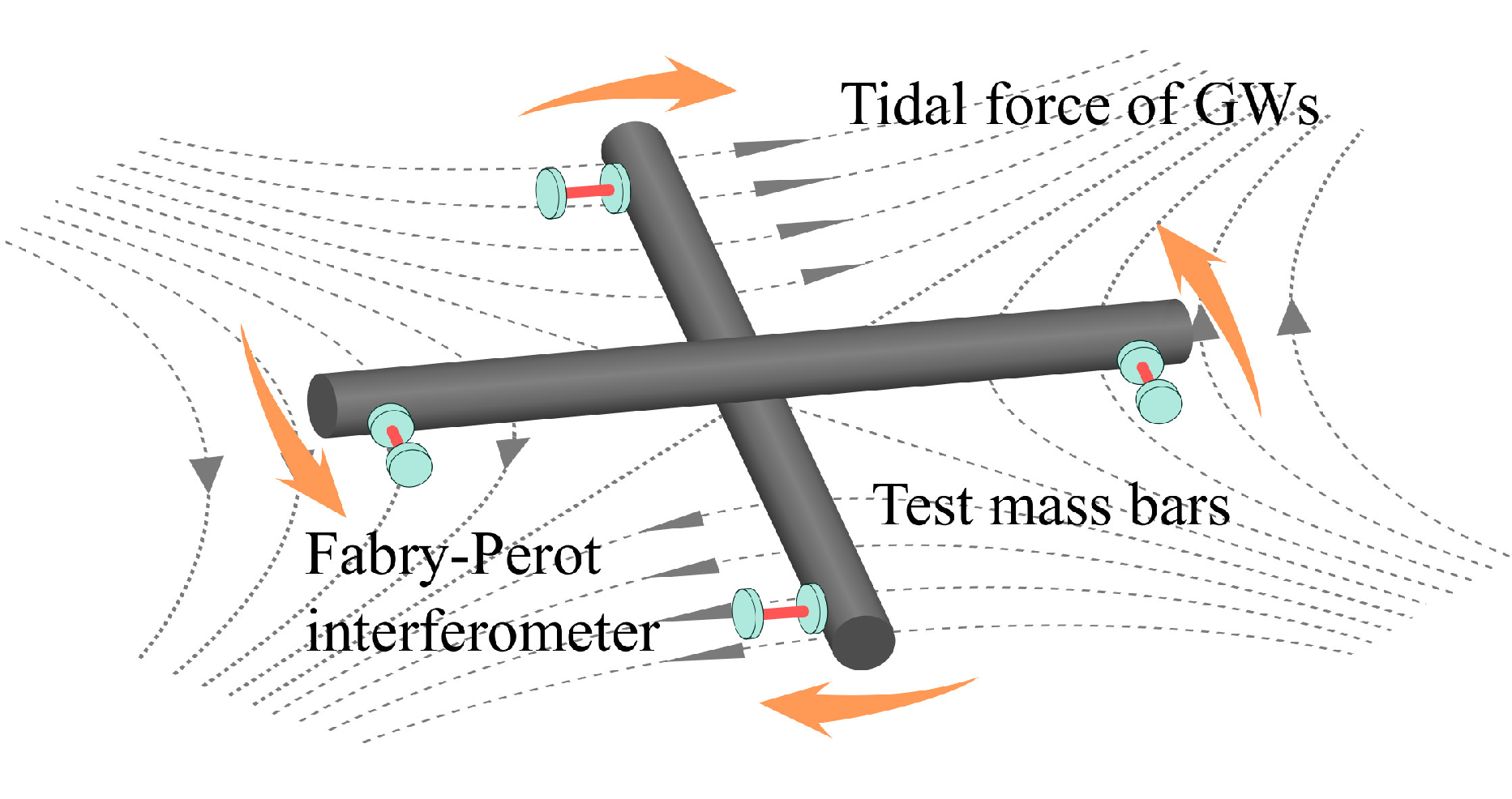}\figsubcap{a}}
	\parbox{4in}{\includegraphics[width=4in]{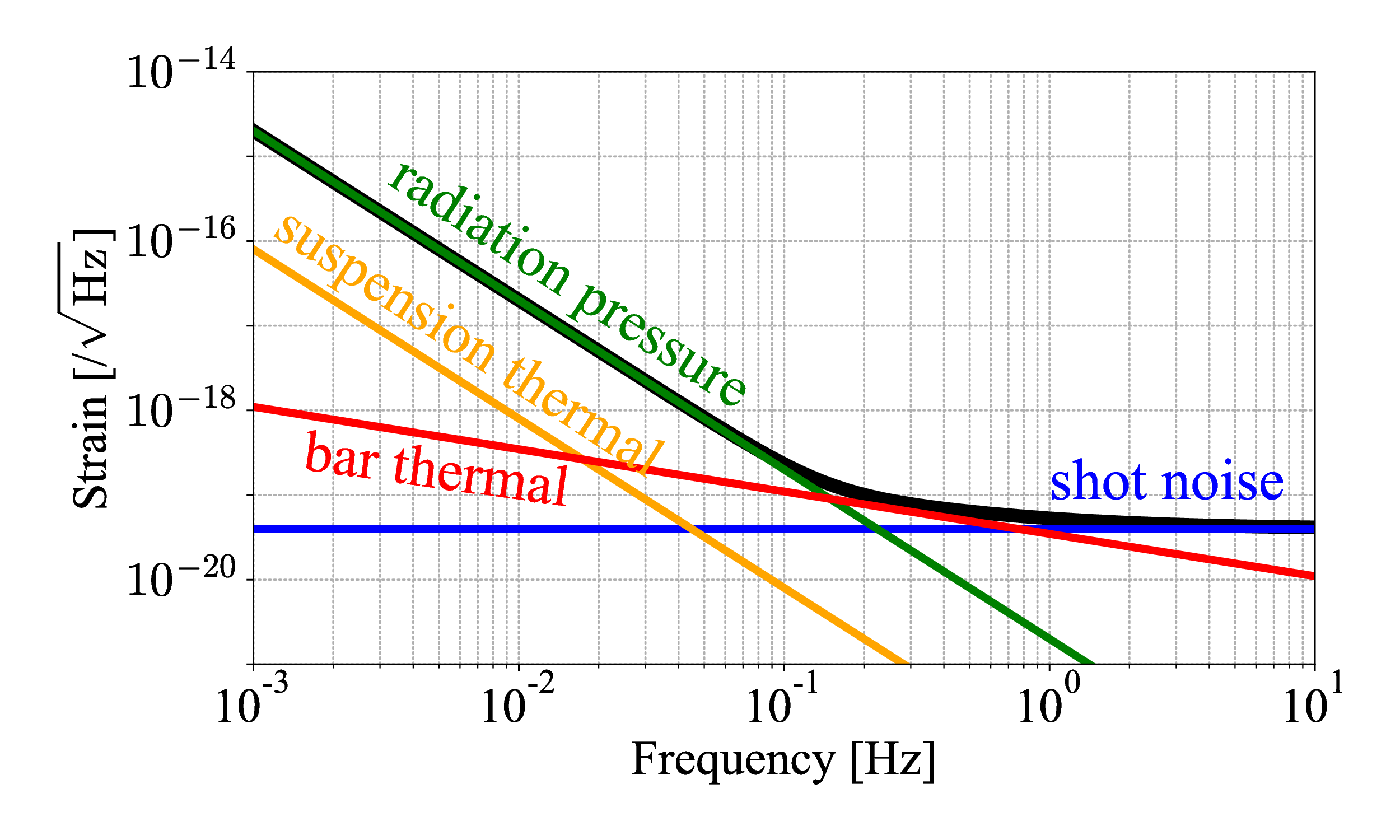}\figsubcap{b}}
	\caption{(a) The working principle of TOBA. Two orthogonal bars rotate differentially when GWs pass. The rotations of the two bars are measured by interferometric sensors. (b) The target sensitivity of TOBA. It is mainly limited by the quantum shot noise and radiation pressure noise. Thermal fluctuations of the bar dominate around 0.3 Hz.}
\label{fig:TOBA}
\end{center}
\end{figure}

The simplified configuration of TOBA is depicted in Fig. \ref{fig:TOBA} (a).
Two orthogonal bars are suspended horizontally by wires (not shown in the figure).
Tidal forces from GWs induce rotations on the bars differentially in proportion to their quadrupole moments, allowing the observation of GWs by measuring their rotations with the interferometric sensors.

\subsection{Gravitational wave observation}
The target sensitivity of TOBA is shown in Fig. \ref{fig:TOBA} (b).
It assumes the use of 10 m long bars at cryogenic temperatures (4 K).
The rotations of the bars are measured with Fabry-Perot cavities at the end of the bars, which have a finesse of 100 and input laser power of 10 W.
The sensitivity is expected to be $10^{-19} /\sqrt{\rm Hz}$ around 0.1 Hz, which is limited by quantum noises (shot noise and radiation pressure noise) and thermal noise of the bar \cite{TOBA0}.

The observable range for IMBHs of $10^5$ $M_{\rm sun}$ reaches 10 Gpc at the target sensitivity \cite{TOBA0}.
The observations of IMBHs (or lack thereof) will provide us important information about the formation process of supermassive black holes, since IMBH coalescence is one of the possible scenarios.

Another important target is the stochastic GW background from the inflationary universe.
In terms of the dimensionless energy density of GWs, $\Omega_{\rm GW}$, the design sensitivity of TOBA corresponds to $\Omega_{\rm GW}\simeq10^{-7}$ with a one-year observation period \cite{TOBA0}.
This is better than the current limit on $\Omega_{\rm GW}$, which was set by an indirect approach from big-bang nucleosynthesis \cite{Maggiore}.

\subsection{Geophysical Applications}\label{sec:geophysics}
Since GW detectors actually measure local gravity gradients, their signals also originate from terrestrial sources such as the ground and the atmosphere.
It was recently proposed that measuring the terrestrial gravity fluctuation can be useful for geophysical purposes, even with a moderate sensitivity of around $10^{-15}$ $/\sqrt{\rm Hz}$ at 0.1 Hz.
Two relevant cases are introduced here.
For these purposes, a similar torsion-bar detector named TorPeDO \cite{TorPeDO} is also under development.

\subsubsection{Earthquake early warning}
After a fault rupture starts, the density of the surrounding ground is gradually redistributed as the seismic waves propagate outward, resulting in transient gravity change.
This transient signal was found to be useful for earthquake early warning \cite{EEW_JH}, which can be earlier than p-wave arrival since gravity propagates at the speed of light.
Such signals were actually detected using a superconducting gravimeter and several broadband seismometers during the Tohoku-Oki earthquake ($M_{\rm w}=9.1$) \cite{EEW_JPM2016,EEW_MV2017}.
For smaller earthquakes whose magnitude are less than about 8, gravity gradiometers such as TOBA are necessary to filter out the effect of seismic vibration.
With the sensitivity of $10^{-15}$ $/\sqrt{\rm Hz}$ at 0.1 Hz, earthquakes of $M_{\rm w}=7.0$ are estimated to be detectable from a distance of 100 km within 10 s.
This sensitivity is feasible with a smaller ($<1$ m scale) TOBA.

\subsubsection{Newtonian noise}
Newtonian gravity fluctuations caused by the moving ground or the atmosphere is called Newtonian noise (NN) \cite{NN_infra,NN_seis}.
NN is estimated to be significant at below 30 Hz, hence the mitigation of it is essential for the future ground-based GW detectors.
In order to reduce NN, feedforward subtraction using noise models and environmental sensors is proposed, but the noise models are not perfect for some types of NN \cite{NN_infra}.
By measuring the NN directly at around 0.1 Hz where the noise amplitudes grows to the order of $10^{-15}$ $/\sqrt{\rm Hz}$, TOBA can test the noise model of NN and its cancellation scheme, which will provide realistic estimation of achievable cancellation.

\section{Prototype developments so far}\label{sec:prototype}
\begin{figure}[t]
\begin{center}
\includegraphics[width=110mm]{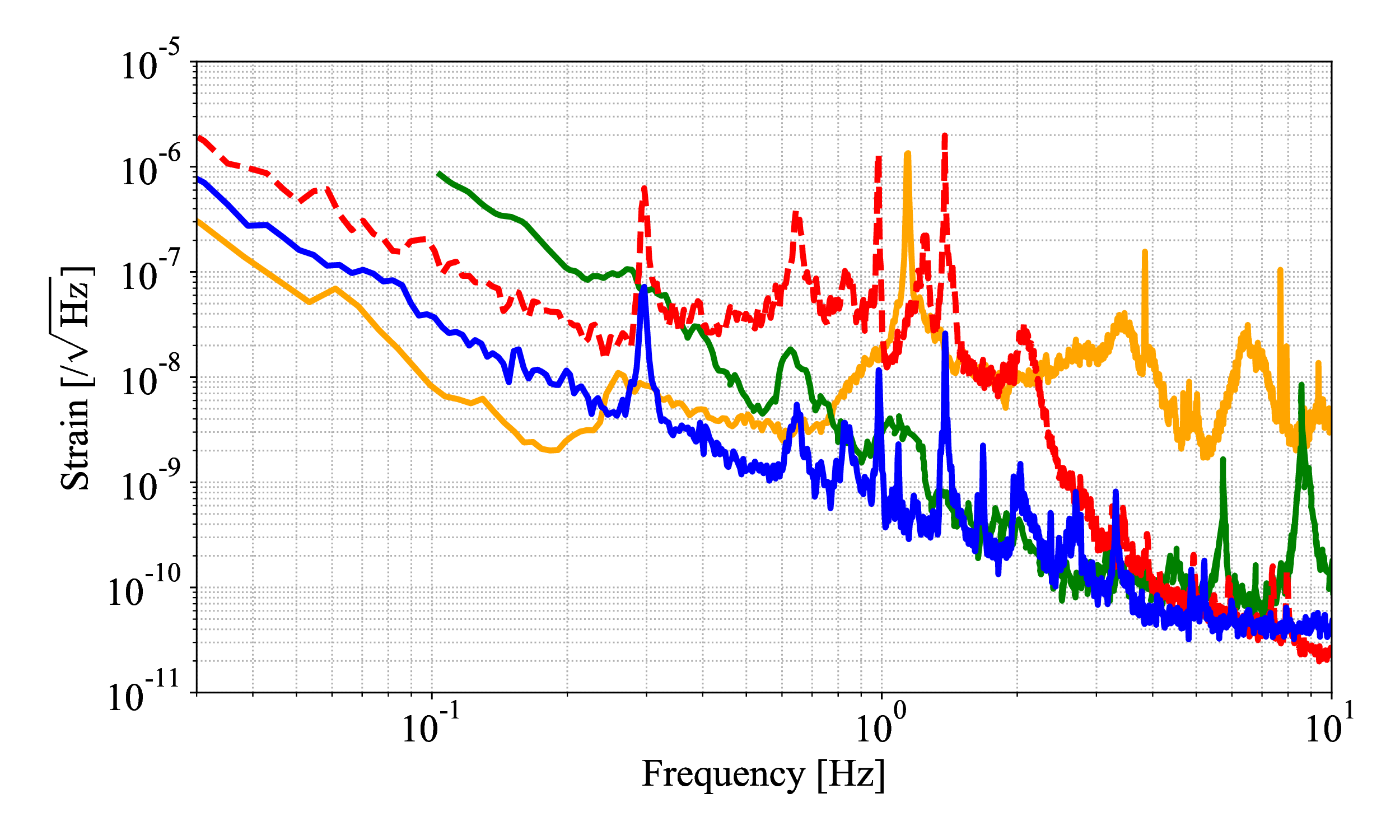}
\end{center}
\caption{The achieved sensitivities of the prototypes: the first prototype (orange) and the second prototype (green). The demonstration result of the seismic cross-coupling noise reduction with a wire-suspended 20 cm bar is shown with dashed red line (before) and solid blue line (after).}
\label{fig:prototype}
\end{figure}

In order to reach the target sensitivity, demonstration of noise reduction is essential before constructing the 10 m scale detector.
Some prototypes of TOBA have been developed so far.
Fig. \ref{fig:prototype} shows the achieved sensitivities of the prototypes.

\subsection{First prototype (phase-I)}
As proof of concept, the first prototype was developed in 2011 with a single 20 cm bar \cite{TOBA_phase1}.
The bar was magnetically levitated in order to make it soft in horizontal rotation.
The sensitivity was $10^{-8}$ $/\sqrt{\rm Hz}$ at 0.1 Hz, which was limited by the magnetic field fluctuation at low frequencies due to the large magnet attached for levitation.
Above 0.1 Hz, cross-coupling noise from translational seismic vibration was significant.

\subsection{Second prototype (phase-II)}
The suspension system was changed to wires in the second prototype \cite{TOBA_phase2}.
The rotations of the bars were measured from the suspended bench which does not respond to GWs.
The vertical rotations were also measured in order to compensate for the blind direction of the horizontal rotations (multi-output configuration).
The noise level equivalent to GWs was $10^{-10}$ $/\sqrt{\rm Hz}$ at 3 Hz, which is thought to be induced by the vibration of the fiber optics for the interferometric sensors.
As with the first prototype, seismic cross-coupling noise was significant around 1 Hz.

\subsection{Seismic cross-coupling noise reduction}
These prototypes succeeded in the identification of many noise sources as shown above.
Among them, the seismic cross-coupling noise was one of the most important noises to reduce since it was not well understood.
In order to clarify how the translation of the ground is transferred to the rotational signal, theoretical investigation and experimental demonstration using a wire-suspended 20 cm bar were performed \cite{TOBA_scc}.
The cross-coupling transfer functions were suppressed down to the order of $10^{-5}$ rad/m, with one of the cross-coupling routes reaching $5\times10^{-7}$ rad/m.
Although these values are not sufficient for future sensitivity, the basic reduction scheme had been successfully demonstrated.

\section{Next step : phase-III TOBA}\label{sec:phase3}
Following the previous prototypes, an upgraded configuration is currently under development (phase-III).
The main purpose is to complete the demonstration of noise reduction and reach a sensitivity of about $10^{-15}$ $/\sqrt{\rm Hz}$ at 0.1 Hz.
In addition to the noises found in the prototypes, reducing thermal noise is an important target.

The configuration is shown in Fig. \ref{fig:phase3} (a).
Two 35 cm dumbbell-shaped bars will be suspended from the intermediate mass, and the whole torsion pendulum system is connected to the active vibration isolation table.
The horizontal rotation of each bar is measured from the optical bench, and the differential signal between the bars is used for observation.
Hence the common rotation signal, such as the rotations of the intermediate mass or the optical bench, can be largely canceled.
The target sensitivity with this configuration is shown in Fig. \ref{fig:phase3} (b). 

\def\figsubcap#1{\par\noindent\centering\footnotesize(#1)}
\begin{figure}[t]%
\begin{center}
	\parbox{2.8in}{\includegraphics[width=2.8in]{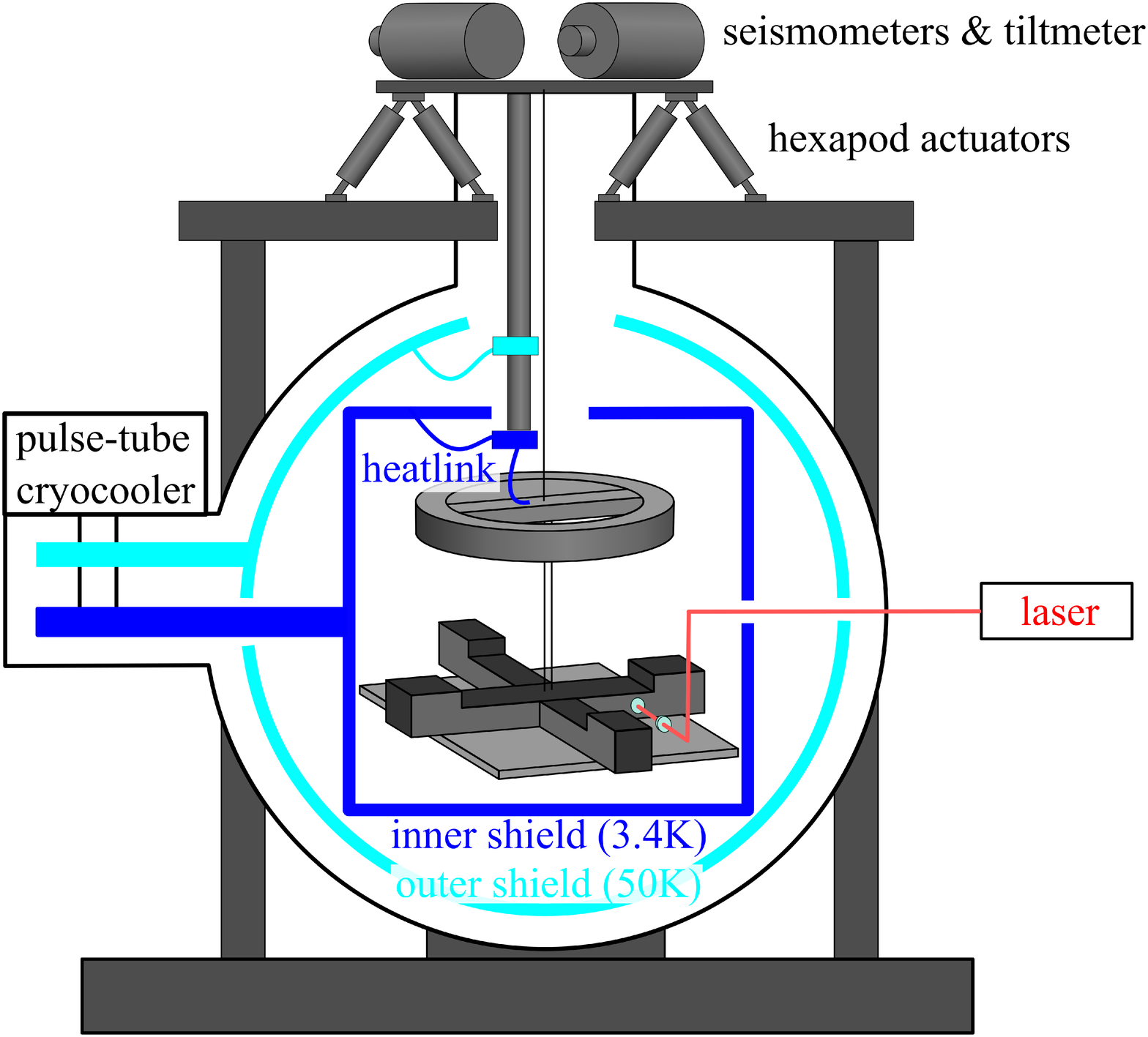}\figsubcap{a}}
	\parbox{4in}{\includegraphics[width=4in]{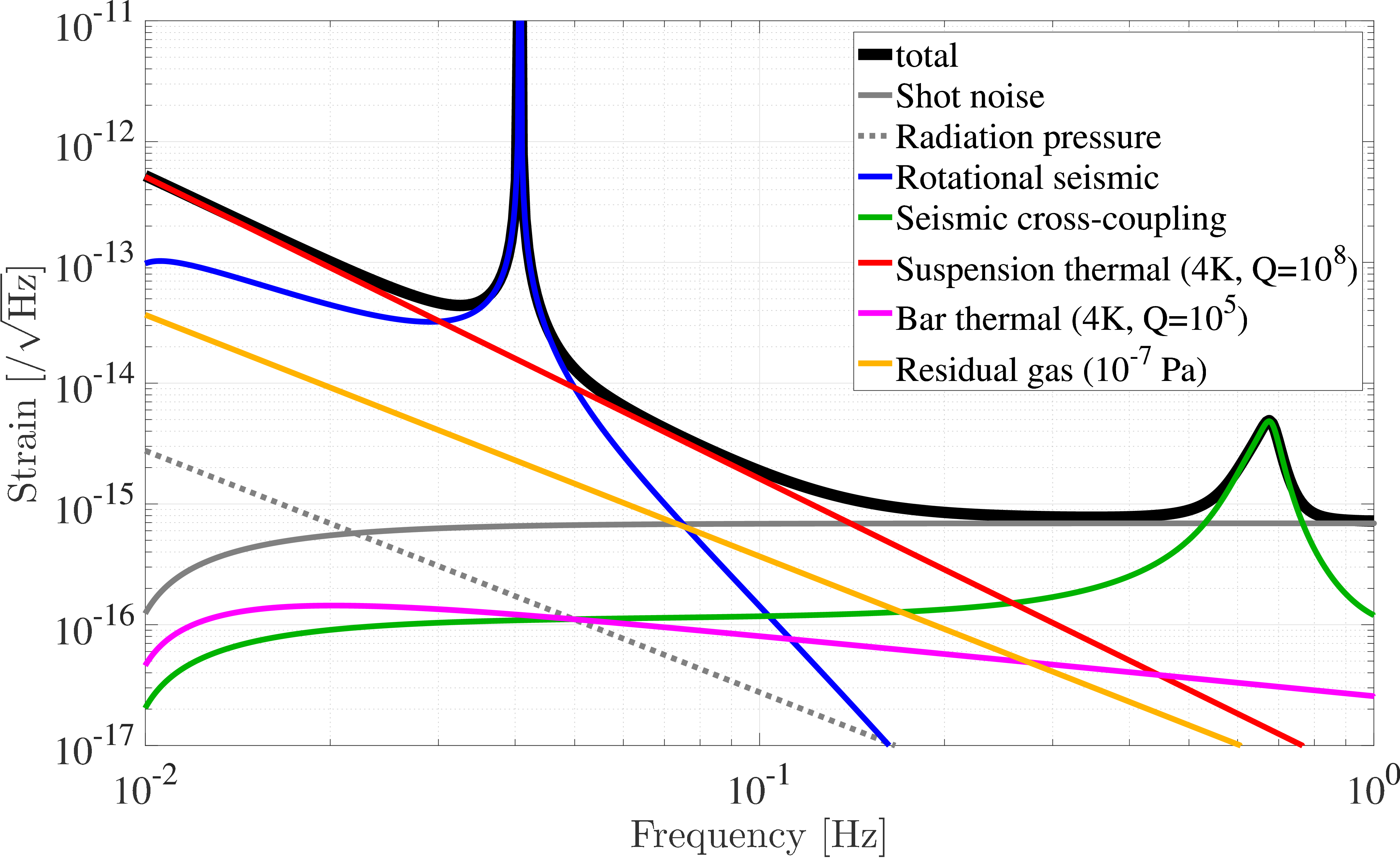}\figsubcap{b}}
	\caption{(a) The design of phase-III TOBA. (b) The target sensitivity of phase-III TOBA.}
\label{fig:phase3}
\end{center}
\end{figure}

The rotations of the bars are measured with Fabry-Perot cavities.
The TEM$_{10}$ mode of the Gaussian laser beam generated by the rotation of the bar is amplified by the cavity, and detected as the rotational signal.
The cavities are designed to have a finesse of 300, and the input power of the laser is 50 mW.

The suspended pendulums will be kept at the cryogenic temperature to reduce thermal fluctuations of the suspension wires and the bars, whose contributions to the sensitivity are depicted in Fig. \ref{fig:phase3} (b) with the red line and the pink line, respectively.
The pulse-tube cryocooler cools the inner radiation shield to about 3.4 K, and it is connected to the intermediate mass via pure aluminum wires (heatlinks).
Because of the heat inflow from the top stage, the equilibrium temperatures of the intermediate mass and the test masses are 4.0 K.
The heating from the input laser is small enough, whose effect on the temperature is estimated to be less than 0.1 K. 
The test masses are cooled via thermal radiation from the surface (300 K $\rightarrow$ 50 K) and heat conduction of the suspension wire (50 K $\rightarrow$ 4 K).
The suspension wires will be made of silicon or sapphire, which are known to have high Q values ($\sim 10^8$) at cryogenic temperatures \cite{siliconQ}.
The bars are made of copper, whose Q value is assumed to be $10^5$.
These high Q values contribute to the reduction of thermal fluctuations.

Additionally, the residual gas around the bars, which is one of the sources of torque noise, is reduced to $10^{-7}$ Pa by the cryopump effect of the shields. 
This reduces the noise below $10^{-15}$ $/\sqrt{\rm Hz}$ at 0.1 Hz, as shown in Fig. \ref{fig:phase3} (b) with the orange line.

The six seismometers and the hexapod actuators are placed at the top stage to keep it isolated from the ground vibration by active feedback control. 
An additional ground tiltmeter is located on the stage in order to resolve the tilt-horizontal coupling.
The target vibration level is below $10^{-7}$ $\rm{m}/\sqrt{\rm Hz}$ at 0.1 Hz, which is required to reduce seismic cross-coupling noise sufficiently.
It is also important for the stable operation of the system as it would reduce the RMS amplitude of the pendulum vibration.
The rotational vibration of the ground is kept below $10^{-10}$ $\rm{rad}/\sqrt{\rm Hz}$ at 0.1 Hz with the actuators, by using the common rotation of the two bars.
The noise from the translational and the rotational vibration of the ground are shown in Fig. \ref{fig:phase3} (b) with the green line and the blue line, respectively.

By combining these systems, the sensitivity reaches about $10^{-15}$ $/\sqrt{\rm Hz}$ at 0.1 Hz, as shown in Fig. \ref{fig:phase3} (b).
As already stated in Sec. \ref{sec:target}, this sensitivity will enable us to detect earthquakes and Newtonian noise.
Therefore, phase-III TOBA is an important step for both the geophysical application as well as technical demonstration.

\section{Summary}\label{sec:summary}
An overview of TOBA has been provided, with the current status being that of technical demonstration.
A small scale prototype with a cryogenic suspension system is under development, targeting a sensitivity of $10^{-15}$ $/\sqrt{\rm Hz}$.
If the development goes according to plan, geophysical targets such as earthquakes and Newtonian noise are expected to be detectable.
After that, we will move on to increasing the size of the detector to achieve the final target sensitivity of $10^{-19}$ $/\sqrt{\rm Hz}$, which would enable us to observe gravitational waves.

\section*{Acknowledgments}
The work about TOBA is supported by JSPS KAKENHI Grants Number JP16H03972, JP24244031 and JP18684005.

\end{document}